\documentclass[times, review, 10pt]{elsarticle}
\usepackage{amssymb}
\usepackage{amsmath}
\usepackage{graphicx}
\usepackage{booktabs}
\usepackage{multirow}
\usepackage{xcolor,colortbl}
\usepackage{tocloft}
\usepackage[ruled,vlined]{algorithm2e}
\usepackage{pifont}
\usepackage{hyperref}
\usepackage{caption}
\usepackage{subcaption}

\journal{Pattern Recognition}

\begin{document}

\begin{frontmatter}

\title{Cell as Point: One-Stage Framework for Efficient Cell Tracking}

\author[label1]{Yaxuan Song}
\author[label1]{Jianan Fan\corref{cor1}}
\author[label2]{Heng Huang}
\author[label3]{Mei Chen}
\author[label1]{Weidong Cai\corref{cor1}}

\cortext[cor1]{Corresponding authors.}

\affiliation[label1]{organization={School of Computer Science},
            addressline={The University of Sydney},
            city={Sydney},
            postcode={2006},
            state={NSW},
            country={Australia}}

\affiliation[label2]{organization={Department of Computer Science},
            addressline={University of Maryland College Park},
            city={College Park},
            postcode={20742},
            state={Maryland},
            country={United States}}
            
\affiliation[label3]{organization={Microsoft},
            city={Redmond},
            postcode={98052},
            state={Washington},
            country={United States}}

\begin{abstract}
Conventional multi-stage cell tracking approaches rely heavily on detection or segmentation in each frame as a prerequisite, requiring substantial resources for high-quality segmentation masks and increasing the overall prediction time.
To address these limitations, we propose \textbf{CAP}, a novel end-to-end one-stage framework that reimagines cell tracking by treating \textbf{C}ell \textbf{a}s \textbf{P}oint. 
Unlike traditional methods, CAP eliminates the need for explicit detection or segmentation, instead jointly tracking cells for sequences in \textbf{one stage} by leveraging the inherent correlations among their trajectories. 
This simplification reduces both labeling requirements and pipeline complexity.
However, directly processing the entire sequence in one stage poses challenges related to data imbalance in capturing cell division events and long sequence inference.
To solve these challenges, CAP introduces two key innovations: (1) adaptive event-guided (AEG) sampling, which prioritizes cell division events to mitigate the occurrence imbalance of cell events, and (2) the rolling-as-window (RAW) inference strategy, which ensures continuous and stable tracking of newly emerging cells over extended sequences.
By removing the dependency on segmentation-based preprocessing while addressing the challenges of imbalanced occurrence of cell events and long-sequence tracking, CAP demonstrates promising cell tracking performance and is \textbf{8 to 32 times more efficient} than existing methods. 
The code and model checkpoints are available at \href{https://github.com/YXSong000/CAP}{https://github.com/YXSong000/CAP}.
\end{abstract}

\begin{keyword}
Cell tracking \sep Point tracking \sep RAFT \sep End-to-end one-stage framework
\end{keyword}

\end{frontmatter}

\section{Introduction}
\label{sec:intro}

Cells are the fundamental units of organs in the human body~\cite{emami2021computerized}.
Cell events are diverse, continuously changing, and intricate, involving cell migration (translocation), mitosis (division), and apoptosis (death)~\cite{li2008cell}.
Cell tracking is the task of keeping track of each cell's location and reconstructing cell lineages (mother-daughter relations), which help develop preventive therapies~\cite{rodriguez2012cell} and drug discovery~\cite{eccles2005cell}.
Traditional methods of manual tracking demand significant resources and expertise to achieve reliable results. 
These challenges highlight the necessity for developing efficient and automated cell tracking systems.

Several existing works and platforms~\cite{gomez2024celltrackscolab,gallusser2024trackastra} have demonstrated effective performance in achieving automated cell tracking. 
The taxonomies of strategies typically involve segmentation followed by linking (SegLnk)~\cite{moen2019accurate}, segmentation and linking (Seg-\&Lnk)~\cite{embedTrack}, or even detection prior to the segmentation and linking stages (DetSegLnk~\cite{pena2020j} or DetLnkSeg~\cite{li2008cell})~\cite{mavska2023cell}.
Also, frameworks~\cite{Hayashida2020MPM,Hayashida_2022_WACV} often incorporate isolated data preprocessing stages to represent cell tracking data.
These multi-stage cell tracking pipelines can limit the overall performance of the framework, constrained by the errors that occur at earlier stages and are accumulated and amplified at subsequent stages~\cite{Li_2022_CVPR}.
Additionally, a multi-stage framework involving both detection/segmentation and tracking not only requires the tracking ground truth but also demands additional labels, i.e., high-quality segmentation masks, to train the model.
It is highly labor-intensive and time-consuming to produce the data annotation, rendering such methods impractical for clinical applications.

\begin{figure}
\centering
\centerline{\includegraphics[width=0.7\textwidth]{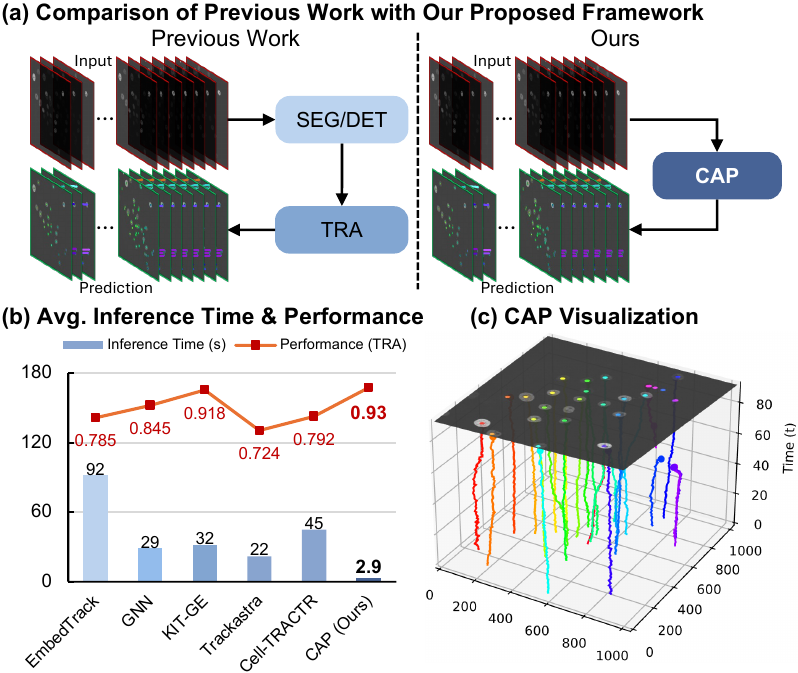}}
\caption{Our proposed \textbf{CAP} is an end-to-end trained framework leveraging the idea of \textit{Cell as Point} to facilitate tracking cells efficiently. 
As (a) illustrates, instead of previous work requiring segmentation (SEG) or detection (DET) as a prerequisite for final tracking (TRA), CAP is able to track all cells within sequence frames in \textbf{one stage}.
(b) shows that CAP reduces the inference time ($2.9$s) by approximately \textbf{$\mathbf{8}$ to $\mathbf{32}$ times} compared to previous works while maintaining high tracking performance of $0.93$.
(c) demonstrates the cell tracking result predicted by CAP.}
\label{vis_merge}
\end{figure}

To address the limitations of existing multi-stage frameworks, we propose a novel \textbf{one-stage} cell tracking framework, \textbf{Cell as Point} ({abbreviated as} \textbf{CAP}), as demonstrated in Figure~\ref{vis_merge} (a), that treats each cell as a point to jointly track numerous cell points within a sequence.
The one-stage CAP framework not only benefits from the efficient prediction procedure through a straightforward pipeline as shown in Figure~\ref{vis_merge} (b), but also from its low requirement for data annotation, solely relying on the tracking ground truth (coarse masks and cell acyclic graphs) to represent the position and status of each cell.
Specifically, by proposing \textbf{cell point trajectory} and \textbf{visibility} to represent the cell tracking data within a one-stage framework comprehensively, CAP offers an elegant solution to alleviate the issues of the complicated data preprocessing and the dilemma of representation of cell decay.
Another key innovation of CAP is the introduction of \textbf{adaptive event-guided} ({abbreviated as} \textbf{AEG}) sampling method and \textbf{rolling-as-window} ({abbreviated as} \textbf{RAW}) inference approach. 
Specifically, the AEG sampling is a temporal data sampling strategy for cell microscopy imaging sequence data to remedy the imbalance of cell division occurrences in the dataset; RAW is an inference approach to capture and track new cells within a long sequence in an efficient way.
{Our contributions are summarized as follows:}
{
\begin{itemize}
    \item  We propose a novel \textbf{CAP} cell tracking framework that is highly efficient in terms of \textbf{inference time} and \textbf{data annotation requirements}, enabling \textbf{one-stage} tracking that bypasses detection / segmentation to investigate cellular behaviors. 
    \item We address prevalent challenges in the cell sequence data by proposing \textbf{AEG} sampling to resolve \textit{imbalanced cell division occurrence}, \textit{representing complex cell events} for training purposes through \textbf{cell point trajectory} and \textbf{visibility} concepts within a one-stage framework.
    \item We introduce a \textbf{RAW} inference approach in the CAP framework to \textit{accurately capture and track new cells in long sequences} during frame sequences.
    \item Our proposed one-stage approach, CAP, achieves SOTA performance and is 8 to 32 times more efficient for inference than existing multi-stage methods, while alleviating the need for detection or segmentation.
\end{itemize}
}

\section{Related Work}
\label{sec:rel_work}
\subsection{Cell Tracking}

Cell tracking is a crucial task in biological imaging that typically involves the identification and temporal association of cell instances across image frames. 
Traditionally, this process includes two main steps: cell segmentation (detection)~\cite{wolf2023unsupervised} and cell linking~\cite{CHEN2021101}. 
The segmentation step identifies individual cells in each frame, while the linking step establishes temporal associations between these cells across frames, sometimes capturing complex dynamics such as cell division and migration~\cite{embedTrack,mavska2023cell}. 
Nowadays, instead of traditional approaches such as Nearest Neighbors linking~\cite{stegmaier2017fuzzy}, Bayesian filters~\cite{hoseinnezhad2012visual,hirose2017spf}, Viterbi~\cite{6957576,magnusson2015tracking}, and graph-based matching~\cite{arbelle2018probabilistic,turetken2016network,loffler2021graph}, the advent of deep learning has transformed the landscape of cell tracking, leveraging advanced neural architectures for segmentation and tracking tasks.
Deep learning approaches often employ complex models, e.g., CNN, RNN, GNN, and transformer, to combine multiple stages like detection~\cite{zhou2019joint,hayashida2019cell,jung2019multiple,nishimura2020weakly,Hayashida2020MPM,Hayashida_2022_WACV,kirsten2025cell}, segmentation~\cite{7164093,7493415,embedTrack,bao2021dmnet,ben2022graph,scherr2020cell,gallusser2024trackastra,o2025cell}, and tracking into the frameworks. 
Contemporary methods~\cite{rempfler2017cell,payer2018instance,hayashida2019cell} aim to integrate contextual cell data into tracking models.
ConvGRU~\cite{payer2018instance} is introduced by Payer et al., capturing local cellular and inter-frame features.
However, this method requires extensive annotated training data for all cell regions, making it labor-intensive.
Motion and Position Map (MPM) is capable of representing detection and association at once, which is proposed by Hayashida et al. \cite{Hayashida2020MPM}, marking an advancement in simplifying the pipeline for cell tracking. 
However, it still requires multiple stages, first generating the MPM followed by subsequent detection and tracking procedures.

Thus, although existing multi-stage methods, which are based on individual cell detection followed by linking / association or evolving segmentation masks over time, show great cell tracking performance, they heavily rely on segmentation as a preliminary step before tracking.
These methods often require high-quality data annotation for training and can be computationally intensive, especially when dealing with high-resolution data~\cite{yang2023deep}. 
Additionally, these methods often involve complex pipelines that may not support end-to-end or one-stage training, posing challenges for both training and inference phases in terms of efficiency and scalability~\cite{Li_2022_CVPR}.
Thus, to tackle these issues, an end-to-end one-stage framework that omits the intermediate segmentation or detection stage is essential.

\subsection{Point Tracking}
In the field of natural images, recent end-to-end video point-tracking frameworks have achieved commendable performance.
Notably, TAP-Vid~\cite{doersch2022tap} pioneered tracking any point across video frames, establishing a benchmark, and claiming a baseline method. 
However, this method does not support tracking occluded points, limiting its applicability in cell-tracking scenarios. 
Even though PIPs~\cite{harley2022particle} and PointOdyssey~\cite{zheng2023point} methods take measures to address the issue of point occlusion in long sequences by tracking selected points using a fixed sliding window and resuming tracking from the last visible frame, these methods struggle with point tracking independently, which may not capture the dependencies between multiple points. 
OmniMotion~\cite{wang2023omnimotion} accounts for estimated correspondences within a canonical space by optimizing a volumetric representation for video sequences. Although innovative, this method involves test-time optimization, making it computationally expensive and less viable for clinical applications.
CoTracker~\cite{karaev2023cotracker} represents significant advancements in the field of point tracking.
It addresses the inaccuracies led by independent tracking among points by incorporating incremental point coordinates updates based on 4D correlation volumes, proposed by RAFT~\cite{teed2020raft} constructed for all pairs of points from point/object features.
CoTracker3~\cite{karaev2024cotracker3} is advanced by involving a semi-supervised training methodology to reduce reliance on synthetic datasets and utilizing pseudo-labeling techniques to generate training labels from unannotated real-world videos.

However, the trackers designed for general-purpose applications are ineffective for cell tracking because the unique properties of cell behaviors, i.e., cell mitosis and apoptosis, have not been considered in the model construction.  
Specifically, cells are not only relative with position and occlusion but also potentially associated with their ancestors due to the regulation of cellular life decay and division.
Therefore, they cannot simultaneously represent or predict both cell positions and division states with their ancestors or children.
Moreover, regarding the publicly available cell tracking data~\cite{moen2019accurate,mavska2023cell,ulman2017objective,ker2018phase,anjum2020ctmc}, the diversity of cell types results in significant variations in cell division rates and sequence length across different datasets categorized by specimens. 
Imbalanced cell event occurrence~\cite{krawczyk2016learning} in a sequence can lead to the omission of complete division events (from the death of the mother cell to the birth of the daughter cell) during random sampling with a low division rate; the inference for long sequences typically requires more time or computational resources to predict cell locations and lineages at once.
Thus, existing methods potentially result in the \textit{under-learning of division cases} in the model training process and a higher resource requirement in the \textit{long-sequence inference} procedure. 
Thus, the uniqueness of this time-lapse cell sequential image data poses the most significant challenge in developing a one-stage cell tracking framework to effectively track cells jointly.

\section{Methodology}

\label{sec:method}

\subsection{Preliminaries}

\label{preliminary}

Recurrent All-Pairs Field Transforms (RAFT)~\cite{teed2020raft} embedded in CoTracker~\cite{karaev2023cotracker} can be conceptualized as incrementally computing the similarity between all pairs of pixel points across frames in the sequence, enabling the model to handle large displacements and complex motion patterns.
In the field of cell tracking, this design promotes the cell point's trajectories in a frame sequence constructed by iteratively updating the flow field using a recurrent unit that performs lookups on a 4D correlation volume.

\paragraph{4D Correlation Volume}
Given two feature maps $F_1 \in \mathbb{R}^{H\times W\times dim}$ and $F_2 \in \mathbb{R}^{H\times W\times dim}$ with output $dim$-dimensions, which are features extracted from two input images $I_1$ and $I_2$, by the feature extractor $g_\phi$: $F_1=g_\phi(I_1)$ and $F_2=g_\phi(I_2)$.
RAFT computes a correlation volume $C$ that captures the similarity between all pairs of pixels across the two images. The correlation between a pixel at position $(i,j)$ in $F_1$ and a pixel at position $(k,l)$ in $F_2$ is calculated as the dot product of their feature vectors:
\begin{equation}
    \begin{aligned}
    C_{ijkl} & = \langle F_1(i,j), F_2(k,l) \rangle =\sum_h F_1(i,j,h) \cdot F_2(k,l,h),
    \end{aligned}
\end{equation}
which is a 4D tensor of dimensions $H \times W \times H \times W$, where $H$ and $W$ are respectively the height and width of feature maps $F$s.
To capture multi-scale feature information for large displacement, RAFT constructs a correlation pyramid by applying average pooling to the last two dimensions of the 4D correlation volume.
Specifically, this pooling is performed with kernel sizes $k$ of $1$, $2$, $4$, and $8$, and a set of multi-scale correlation volumes $C^k = \{ C^1, C^2, C^3, C^4 \}$ is generated.
Thus, each volume $C^k$ in the pyramid has dimensions $H \times W \times \frac{H}{2^k} \times \frac{W}{2^k}$.

\paragraph{Iterative Updates}
The update operator estimates a sequence of flow estimations $\{ f_1, \dots, f_n \}$ with initializing the start point $f_0=0$.
For each iteration, it produces $\Delta f$ as the direction that should be applied to the current estimation: $f_{k+1}=\Delta f + f_{k+1}$.
Specifically, given the current flow estimation $f_k$, the correlation volume pyramid is retrieved and undergoes processing through two convolutional layers to extract relevant information.
The updated information is comprised of correlation, flow, and context features and is updated over time as a feature map concatenation.

\subsection{Proposed Method: Cell as Point (CAP)}

\begin{figure}
  \centering
  \centerline{\includegraphics[width=\textwidth]{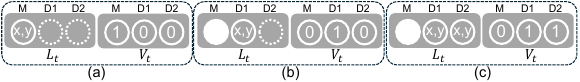}}
\caption{\textbf{Cell Point Trajectory and Visibility.} M, D1, and D2 denote the mother cell, the daughter cell \#1, and daughter cell \#2, respectively. (x,y) in $L_t$ represent the location coordinates, and $0$ or $1$ in $V_t$ respectively represent non-existing and existing cells. Three valid trajectory and visibility status: (a) M has not divided; (b) M has divided, and only D1 occurs in the $t$-th frame; (c) M has divided, and both D1 and D2 occur in the $t$-th frame.}
\label{fig:cell_represent}
\end{figure}

Cell tracking aims to identify the characteristics of cell migration and division, thereby fully automating the prediction of specific cell position and division status changes in each frame of a time-lapse microscopy video.
In our framework, CAP, every single cell is a point, tracking all cell points throughout the duration of a frame sequence $S = (I_t)_{t=1}^T$ containing $T$ grayscale cell microscopy images $I_t \in \mathbb{R}^{H \times W}$.
\textit{Cell point trajectories} and \textit{visibilities} are proposed to be utilized in the CAP framework to represent and predict cell locations and lineages. 
The \textit{cell point trajectory} is constituted by the $N$ cell centroid point locations $L^i_t = (x^i_t, y^i_t) \in \mathbb{R}^{2}$, $i=1,\dots,N$, $t =1,\dots,T$, where $t$ is the time of a frame sequence. 
The \textit{cell point visibility} of the cell $V^i_t \in \{0,1\}$ indicates whether the $i$-th cell point still exists in the frame sequence at time $t$; in other words, if the cell is non-existing or born due to mitosis and apoptosis.
Since any cell division that happened in the event would have been that of the parent cell into two daughter cells~\cite{inoue1981cell}, Figure~\ref{fig:cell_represent} demonstrates the combinations of cell point trajectory and visibility, representing various states of cells, i.e., migration, mitosis, apoptosis, and cell lineages (the mother cell with two potential daughter cells). 
This simple yet effective representation method will efficiently aid the CAP framework in achieving one-stage efficient cell tracking.

\begin{figure*}[t]
  \centering
  \centerline{\includegraphics[width=\textwidth]{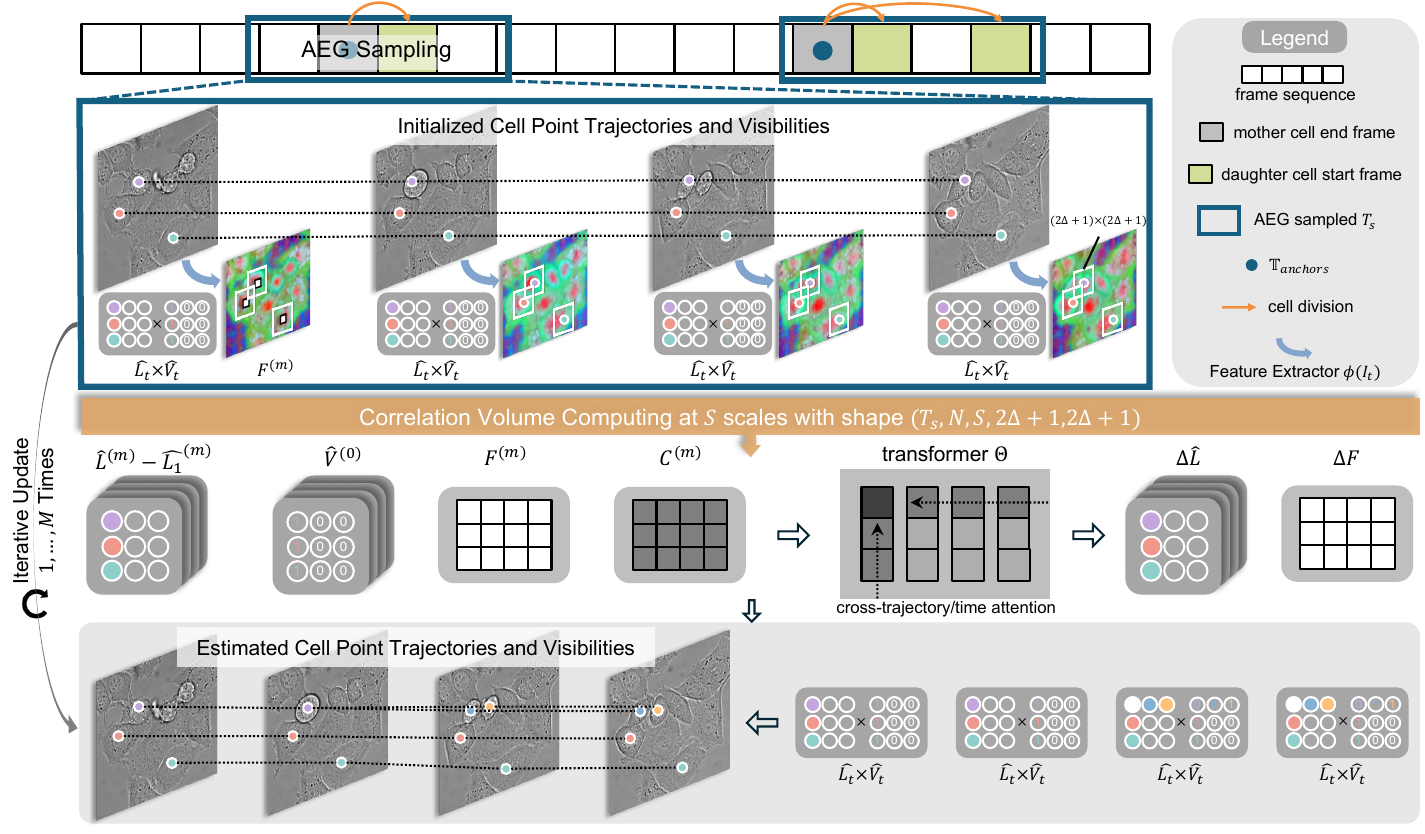}}
\caption{\textbf{Overview of CAP framework.} 
The training sequence $T_s$ is sampled from the entire frame sequence using the AEG strategy. 
For each $T_s$, the cell point trajectories and visibilities are iteratively refined, starting from their initialization. 
For iteration $1,\dots, M$, updates are made to \textit{cell point trajectory} $\hat L$ and \textit{tracking feature} $F$ with \textit{cell point visibility} $\hat V$ computed at the final ($M$-th) update. This figure illustrates a single iteration of the optimization process.
}
\label{fig:main}
\end{figure*}

\subsubsection{Adaptive Event-Guided (AEG) Sampling}
\label{aeg}

To reduce operational costs and compromise overfitting~\cite{babaeizadeh2021fitvid}, instead of seeing the entire training sequence in a single instance, the model learns a continuous sub-sequence of $T_{s}$ frames for each time, where $0 < T_{s} < T$.
However, the imbalanced occurrence of cell division in the entire sequence would cause the model to detect cell division occurring rarely. 
To facilitate the effective training of models to understand the biological behavior of the completed cell division event while mitigating data imbalance~\cite{yudistira2020prediction,liu2021overfitting}, we propose the Adaptive Event-Guided (AEG) sampling technique to balance the occurrences of cell division in a sequence of frames the model sees by determining the initial frame of $T_{s}$ learning a complete event.

The AEG method leverages the intrinsic dynamics of cell division within the frame sequences to guide the sampling process, enhancing the representational diversity of the training dataset and ensuring the model’s robustness to variations in cell behavior.
Technically, as data loading iterates through the whole dataset by detecting the complete cell division processes (from the death of the parent cell to the birth of the daughter cell), an anchor is added at each time cell division occurs $\mathbb{T}_{anchors}$.
The probability of applying AEG $P_\text{AEG}$ is adaptively calculated by $P_\text{AEG}=\frac{N_\text{div} \times T_\text{div}}{T}$, where $N_\text{div}$ denotes the total occurrence of cell division in the entire $T$-frame sequence, and $T_\text{div}$ denotes the duration of the completed cell division.
When the random decimal $p<P_\text{AEG}$ indicates the usage of AEG, our CAP framework incorporates AEG sampling for selecting $T_{s}$, as illustrated at the top of Figure~\ref{fig:main} (dark blue square);
otherwise, the starting frame of $T_{s}$ is randomly located in the $(0,T-T_s)$ range.

For each training sequence, we crop a fixed-length frame sequence in accordance with the starting frame from the video. 
Occasionally, the process of cell division, resulting in the formation of new cells, may span multiple frames.
Thus, the leverage of the AEG approach allows us to capture complete division processes within sampled sequences in data pre-processing. 
This fixed-length cropping is critical to maintaining uniformity in the input data, facilitating more stable and reliable learning outcomes.

\subsubsection{Cell Joint Tracking}

\label{cell_joint}
By regarding each cell as a point and leveraging the two-dimensional feature across time and a set of tracked cell points, we implement an effective method for cell tracking via the correlation between cell point trajectories and visibilities in one stage.
The training process of CAP as a transformer $\Theta$ neural network, where $\Theta: G \mapsto O$.
In order to improve the estimated cell trajectories and visibilities along with $M$ iterations, the trajectories and visibilities of $i$ points (cells) in a $t$-frame sequence are encoded into a grid of input tokens $G^i_t$.
$O^{i}_t$ is the corresponding output of $\Theta$ for representing the refined trajectories and visibilities. 
The following sections elaborate on the transformer formulation visualized in Figure~\ref{fig:main}.

\paragraph{Feature Extraction}
\label{feature_extra}
To initialize the tracking feature $F$ for the part of input tokens $G$, we first generate \textbf{image features} (feature map), $\phi(I_t)$, for each frame in the training sequence, from which the \textbf{tracking features}, $F$, are distilled according to cell point locations.
Specifically, the feature extractor is a convolutional neural network (CNN) trained end-to-end in the framework for computational efficiency.
Leveraging the CNN feature extractor, the feature maps with $dim$ dimensions are generated from $I_t$, $\phi(I_t)\in \mathbb{R}^{dim \times \left(\left\lfloor \frac{H}{k} \right\rfloor \times \left\lfloor \frac{W}{k} \right\rfloor\right)}$, where $k=4$ for computational efficiency, and the features are downscaled by average pooling with strides $s$ for multi-scaled features.
The tracking features $F^i_{t=1}$ for $i$ are captured by bilinear interpolation from the location of cell points on the initial frame $L_{t=1}$ on the feature map $\phi(I_{t=1})$.
The tracking feature vector $F^i_t \in \mathbb{R}^{dim}$ is initialized for the transformer input by broadcasting the tracking features in $t=1$ frame for all $T_s$ frames, which is refined and updated with $M$ iterations.

\paragraph{Correlation Volume}
\label{corr_vol}
To achieve the objective of associating cell point trajectories jointly, CAP employs correlation volume $C^i_t \in \mathbb{R}^S$, where $S=4$, proposed in RAFT~\cite{teed2020raft}.
These correlation volumes $C^i_t$ are derived by assessing the relationship between the tracking features $F^i_t$ and the image features $\phi(I_t)$ in proximity to the estimated cell point location $\hat{L}^i_t$ of the trajectory on the $t$-th frame. 
Technically, each $C^i_t$ is the average over the correlation response in a $2^s\times2^s$ grid:
\begin{equation}
    \begin{aligned}
\relax        [C_t^i]_\delta^s & = \frac{1}{2^{2s}} \sum_\delta^{2^s} \sum_\delta^{2^s} \langle F^i_t, \phi_s(I_t)_{2^s\hat{x}+\delta,2^s\hat{y}+\delta} \rangle \\
        & = \langle F^i_t, \frac{1}{2^{2s}} (\sum_\delta^{2^s} \sum_\delta^{2^s} \phi_s(I_t)_{2^s\hat{x}+\delta,2^s\hat{y}+\delta}) \rangle,
    \end{aligned}
\end{equation}
where $s = 1,\dots, S$ are the feature scales, and $\delta$ is the offset around cell points, $(\hat x, \hat y)$ is the estimated location coordinate of the cell point $\hat L_t^i$.
The value of $[C_t^i]_\delta^s$ is computed as the inner product between the feature vector $F^i_t$ and $\phi_s(I_t)$ at $\hat L_t^i$ pooled with kernel size $2^s \times 2^s$.

\paragraph{Iterative Updates}
In the transformer-based interactive updates, the input tokens express the concatenation of features: location $\hat L^i_t - \hat L^i_1$, visibility $\hat V^i_t$, tracking features $F^i_t$, and correlation volume of cell trajectories $C^i_t$.
As for the output tokens, $O^{i}_t$ contains updates for location and refined tracking features:
$(\Delta \hat L^{3i}_t, \Delta F^{i}_t)$.
CAP estimates \textbf{$3$} locations for each cell, following the representation illustrated in Figure~\ref{fig:cell_represent}: the cell's location and the locations of two potential daughter cells. 
In this way, CAP reserves space for each cell's progeny by representing $L$ and $V$ through a $3$-column format.
$M$-times updates the output tokens by applying the transformer $\Theta$ iteratively for a sequence $T_s$ as shown in Figure~\ref{fig:main}.
Specifically, CAP utilizes transformer $\Theta$ with cross-trajectory/time attention to update both the cell point trajectory, $\hat L$, and the tracking feature, $F$:
\begin{align}
     \hat L^{(m+1)} = \hat L^{(m)} + \Delta \hat L ~\text{and}
     ~ F^{(m+1)} = F^{(m)} + \Delta F, 
\end{align}
where $m=1,\dots,M$, and $m=1$ denotes initialization. 
At the last iteration of the update $M$, cell point visibility $V^{3i}_t$ for locations estimated $\hat L^{3i}_t$ is computed by $\hat V^{(M)} = \sigma(W F^{(M)})$, where $\sigma$ represents the sigmoid activation function and $W$ is a learned matrix of weights.
By investigating the $3$ visibilities $V^{3i}_t$, CAP tells if the cell division occurred on the cell $i$ in the $t$-th frame.

\paragraph{Loss Function}

We supervised the transformer $\Theta$ on the \textit{trajectory} and \textit{visibility} between predicted and ground truth for every semi-overlapping window.
The trajectory loss function $\mathcal{L}_{tra}$ is location regression summed over the iteration updates, and the visibility loss function $\mathcal{L}_{vis}$ is the cross-entropy of binary visibility options:
\begin{align}
    &\mathcal{L}_{tra}(\hat{L}, L) =\sum_{j=1}^{J}\sum_{m=1}^{M}\gamma^{M-m}\|\hat L^{(m,j)} - L^{(j)}\|, \\
    &\mathcal{L}_{vis}(\hat{V}, V)=\sum_{j=1}^{J}\operatorname{CE}(\hat V^{(M,j)},V^{(j)}), 
\end{align}
where $\mathcal{L}_{vis}$ is updated at the $M$-th iteration, and 
$J$ is the total number of $T_s$ whose trajectory starts in the middle of the previous $T_s$. 

\begin{algorithm}
\small
\caption{Algorithm of RAW Inference} 
\label{raw_algo1}
\SetAlgoLined
\textbf{Input:} $\Theta$, $S_\text{infer}\gets (I_t)^{T_\text{infer}}_{t=1}$, $Q\gets (x^i, y^i)$, $l_\text{win}$ \\
\textbf{Output:} $L_t$, $V_t$, where $t\gets1,\dots, T_\text{infer}$\\
\textbf{Initialization}: current process frame $t_\text{cur} \gets 1$ \\
\While{$t_\text{cur}<T_\text{infer}$}{
    Obtain current window frames $S_\text{w} \gets (I_t)^{t_\text{cur}+l_\text{win}}_{t=t_\text{cur}}$\\
    $L_\text{w}$, $V_\text{w} \gets \Theta(S_\text{w})$ \\
    \For{$i \gets 0$ \textbf{to} $l_\text{win}$}{
        $t_\text{cur} \gets t_\text{cur}+i$ \\
        $Q \gets L_{t_\text{cur}}$ \\
        \If{find new cell(s)} {
            Concatenate $Q$ and $\hat L_{t_\text{cur}}^\text{new}$ \\
            \textbf{break} \\
        }
    }
}
\end{algorithm}

\subsubsection{Rolling-as-Window (RAW) Inference}
\label{infer}

In cell tracking, the inference stage is crucial for producing the final tracking mask and establishing cell lineage relationships by analyzing predicted cell locations and division events.
Existing works~\cite{embedTrack,ben2022graph,liu2016automatic} often customize inference and post-processing stages with different configurations for each dataset, adapting to variations in sequence length.
This data-specific tailoring limits their generalizability and broader applicability to other datasets.
In contrast, our proposed Rolling-As-Window (RAW) inference approach in CAP simplifies and unifies these processes while maintaining effectiveness across varying sequence lengths, enabling high-quality inference regardless of computational constraints.
Specifically, a consistent and high-quality inference result is achieved through a rolling loop, which sequentially extracts yet-to-be-inferred clips from the full sequence, processing them within a defined window for inference.

In a certain sequence with $T_\text{infer}$ frames, if the inference window length $l_\text{win}$ supported by either computational resources or architecture is shorter than $T_\text{infer}$, the first rolling of window frames $S_\text{w}=(I_t)^{l_\text{win}}_{t=1}$ is passed into the trained transformer $\Theta$; otherwise, the whole sequence will be inferred at once.
The RAW inference approach is summarized in Algorithm~\ref{raw_algo1}, which will be triggered while the former scenario happens.
Specifically, a $l_\text{win}$-length subsequence starting from $t=1$ is passed into $\Theta$ for prediction. 
The output, cell point trajectories $L_t$ and visibilities $V_t$, where $t=1, \dots, l_\text{win}$, are processed frame-by-frame to update query points $Q$ based on the $L_t$ of identified cells and check for cell division or the emergence of new cells. 
Whenever a new cell appears, the predicted location of this new cell $\hat L_t^\text{new}$ is incorporated into the updated query points $Q$ for the subsequent rolling of analysis by the formula:
\begin{align}
Q_\text{r+1} = \text{concat}(Q_\text{r}, \hat L_t^\text{new}) = \begin{bmatrix} Q_\text{r} \\ \hat L_t^\text{new} \end{bmatrix},
\end{align}
where $r$ is the times of rolling for window inference.
Subsequently, RAW segments a new window sequence starting from the frame where the new cell appeared, preparing it for the next application of $\Theta$.
By adjusting the $l_\text{win}$ based on the capacity of computational resources and $T_\text{infer}$ value, our CAP is capable of performing efficient and high-quality inference on sequences of any length.

\section{Experiments and Results}
\label{sec:exp}

We conducted extensive experiments to demonstrate the high efficacy and efficiency of CAP, our end-to-end, one-stage framework for cell tracking. 
CAP maintains the promising performance, despite requiring minimal inference time and eliminating the need for additional annotations. 
Additionally, we perform comprehensive ablation studies to assess the effectiveness of individual CAP components.

\subsection{Experimental Setup}

\subsubsection{Datasets}
\label{dataset}

We evaluate our proposed framework on DynamicNuclearNet Tracking in DeepCell~\cite{moen2019accurate} public database, HeLa, PC-3, 3T3, and RAW264 datasets divided by specimen, and ISBI cell tracking challenge (ISBI CTC)~\cite{mavska2023cell,ulman2017objective} public datasets, PhC-C2DH-U373 (U373), Fluo-N2DH-GOWT1 (GOWT1), and Fluo-C2DL-Huh7 (Huh7).
These temporal data used for cell tracking, shot by different devices in various durations on diverse specimens, exhibit diverse morphologies and sequence characteristics.

\begin{figure*}
  \centering
  \centerline{\includegraphics[width=\textwidth]{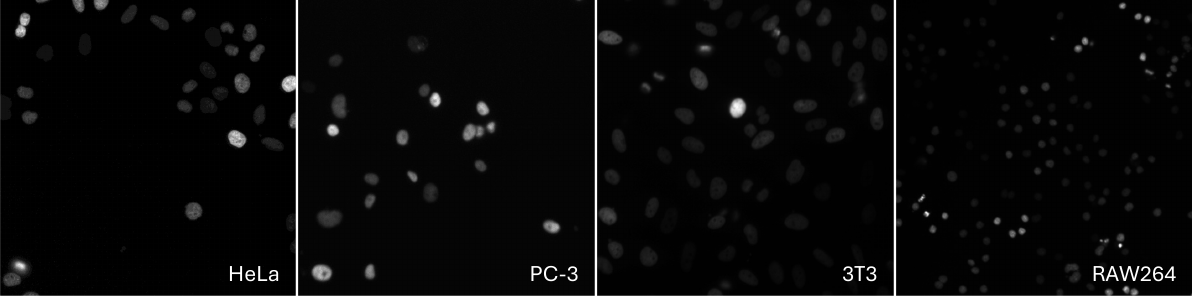}}
\caption{\textbf{Visualization of the DeepCell Database.} The datasets HeLa, PC-3, 3T3, and RAW264 have different cell densities, sizes, and luminances.}
\label{deepcell}
\end{figure*}

\paragraph{DeepCell}
DeepCell database consists of HeLa, PC-3, 3T3, and RAW264 datasets, which respectively contains cell types HeLa-S3, PC-3, NIH-3T3, and RAW 264.7, visualized in Figure~\ref{deepcell}. 
These four cell types are mammalian cell lines that were acquired from the American Type Culture Collection~\cite{moen2019accurate}.
The cells are cultured in Dulbecco’s modified Eagle’s medium (RAW 264.7 and NIH-3T3) or F-12K medium (Hela-S3 and PC-3) supplemented with the necessary drugs, e.g., L-glutamine, penicillin, for all other cells.
DeepCell contains $96$ sequences in total: $45$ for HeLa, $7$ for PC-3, $10$ for 3T3, and $34$ for RAW264. 
Among them, there are $5$, $2$, $1$, and $3$ sequences for datasets HeLa, PC-3, 3T3, and RAW264 for evaluating the training outcomes of the remaining sequences, respectively.
The detailed information on each cell type is illustrated below:

\begin{itemize}
    \item \textit{HeLa} includes $77,180$ objects, $2,234$ tracks, and $189$ cell divisions, with a relatively high cell density. Each sequence is segmented into $42$ frames, lasting $6$ minutes, containing cell behaviors, i.e., cell migration and division. For each frame in the sequences, they are $540\times540$ with a pixel size $0.65$.
    \item \textit{PC-3} is the smallest subset in DynamicNuclearNet, with $5,051$ objects, $159$ tracks, and only $5$ divisions, captured in a $5$-minute duration. 
    All sequences are segmented into $50$ frames with a size $584 \times 584$ and pixel size $0.55$.
    \item \textit{3T3} cells are fibroblasts from mouse embryonic tissue. In this paper, this cell line contributes $62,983$ objects, $1,413$ tracks, and $250$ divisions, offering insights into cell growth and cycle dynamics. 
    There are $71$ frames in each sequence, and each frame is $512\times512$ with a pixel size of $0.65$.
    \item The \textit{RAW264} cell line, derived from murine macrophages, includes $335,049$ objects, $8,642$ tracks, and $1,322$ divisions, making it the largest subset in our study. 
    Each sequence has $45$ frames with a size of $540\times540$ with a pixel size of $0.65$. 
    The cell density in this dataset is the highest, with the smallest cell size.
\end{itemize}

\paragraph{ISBI CTC}

\begin{figure*}[t]
  \centering
  \centerline{\includegraphics[width=0.8\textwidth]{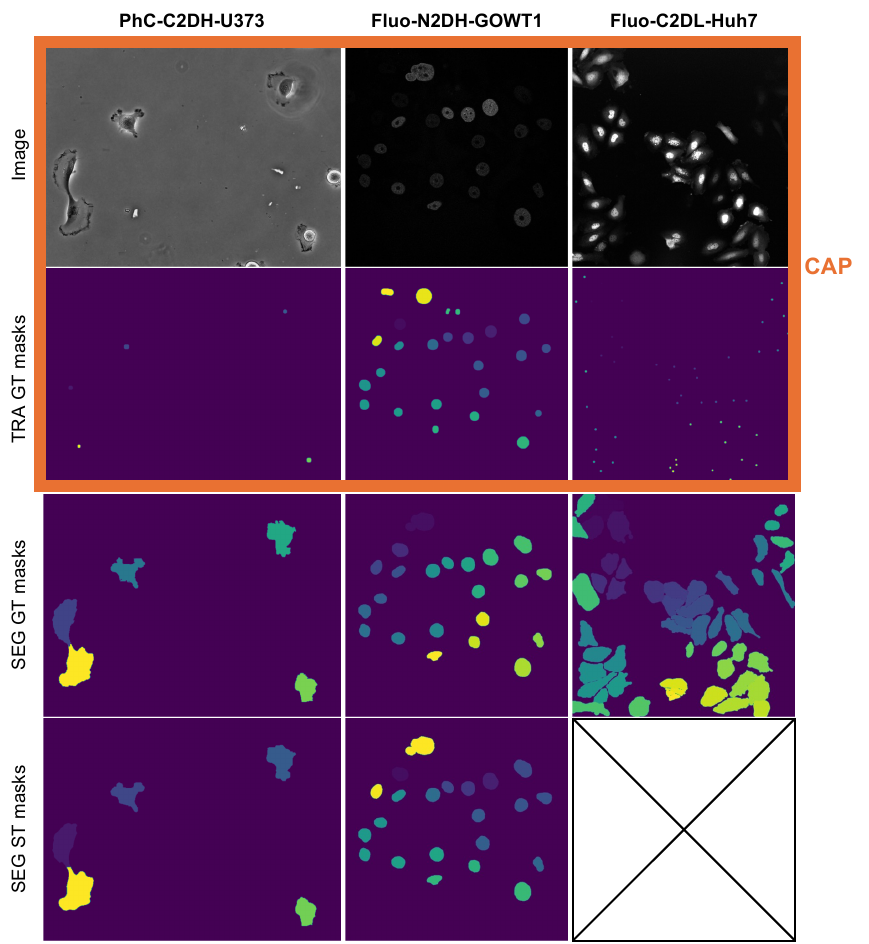}}
\caption{\textbf{Comparison of Quality and Requirement of Data Annotation.} Fluo-C2DL-Huh7 does NOT have segmentation ST. Instead of previous works utilizing all types of mask information with original images in the figure, CAP only utilizes the Tracking GT masks with original images (in the orange block).}
\label{annotation}
\end{figure*}

Figure~\ref{annotation} illustrates datasets we leveraged, U373, GOWT1, and Huh7, with the comparison between tracking (TRA) GT mask, segmentation (SEG) GT, and segmentation (SEG) ST if available.
We observe that SEG GT and ST are much more refined and high-quality than TRA GT masks, requiring more resources and manpower.
In terms of data annotation requirements, unlike previous works that rely on all available TRA GT masks, SEG GT, and ST (when available), CAP requires only the TRA GT masks, thereby substantially reducing the annotation demand.
The detailed information on data collection and the characteristics of each dataset is described below:

\begin{itemize}
\item The \textit{PhC-C2DH-U373} dataset comprises active glioblastoma-astrocytoma U373 cells on a polyacrylamide substrate.
    Each sequence is captured by a Nikon microscope with an object lens Plan Fluor DLL 20x/0.5 for a relatively long duration of 15 minutes, totaling $115$ frames with the size of $520\times696$.
    The cells are relatively sparse in frames.
    The distinct feature of this dataset, compared to others, is its higher grayscale background.
\end{itemize}

\begin{itemize}
    \item Dataset \textit{Fluo-N2DH-GOWT1} documents the activity of GFP-GOWT1 mouse stem cells over a 5-minute period as observed by Leica TCS SP5 microscope with objective lens Plan-Apochromat 63x/1.4 (oil), segmented into $92$ frames with a size of $1024 \times 1024$ per sequence.
    The target cells for tracking are notably active in cell events and exhibit a relatively high density within the frames.
    \item \textit{Fluo-C2DL-Huh7} is a dataset of human hepatocarcinoma-derived cells expressing the fusion protein YFP-TIA-1. 
    Cell events are captured by the Nikon Eclipse Ti2 microscope with a lens CFI Plan Apo Lambda 20x/0.75 in a duration of 15 minutes with $30$ frames for each sequence.
    Cells in this dataset exhibit a relatively low frequency of cell events; However, due to the lower frame rate, cells exhibit significantly larger movements within a sequence.
    Cells are relatively small and have high density in frames. 
\end{itemize}

\subsubsection{Implementation Details}

\begin{table*}[t]
\centering
\resizebox{0.9\textwidth}{!}{%
\setlength{\tabcolsep}{6mm}{
\begin{tabular}{cccc}
\toprule
\textbf{Layer / Block} & \textbf{Kernel} & \textbf{\#Layers / Blocks} & \textbf{Normalization}\\
\midrule
Conv2D & $7\times7$ & 1 & -- \\
Residual blocks & $3\times3$  & 8 & InstanceNorm\\
Conv2D & $3\times3$ & 1 & --  \\
Conv2D & $1\times1$ & 1 & -- \\
\bottomrule
\end{tabular}
}}
\caption{\textbf{Architecture Details of the 2D CNN Feature Extractor.}}
\label{2dcnn}
\end{table*}

For the \textbf{training} process, we utilize $85$ training sequences from the DeepCell database and sequences $02$ from the ISBI dataset, using only tracking (TRA) ground truth (GT). During each iteration, we crop sequences of $T_s = 24$ frames with a sliding window length of $l_\text{slide} = 8$.
We use pre-trained checkpoints from CoTracker~\cite{karaev2023cotracker}, which was trained solely on synthetic RGB data.
Since our input images are grayscale, we replicate the single channel three times to match the required RGB format.
We use an end-to-end trained $2$-dimensional CNN architecture as the encoder to extract tracking features $F$.
Specifically, it integrates bilinear interpolation and utilizes it to extract $S=4$ different scaled image features with $128$-channel output features concatenated to optimized dynamic feature extraction. The architecture details of the 2D CNN feature extractor are listed in Table~\ref{2dcnn}. 
{All the experiments are trained and tested on an NVIDIA GeForce RTX 4090 (24GB) GPU. The total end-to-end training time takes approximately 38 minutes for the DeepCell database and 20 minutes for the CTC datasets. The entire CAP framework contains a total of 194.8M learnable parameters.}
While training the transformer, we follow the pre-trained checkpoint that downsamples the resolution of frames to $384 \times 512$ and choose $S = 4$, $\Delta = 3$ to calculate 4D multi-scaled correlation volumes.
As for the optimization during the training, we present an optimizer configuration in Table~\ref{hyperparameters}.

\begin{table*}[t]
\centering
\resizebox{0.8\textwidth}{!}{%
\setlength{\tabcolsep}{3mm}{
\begin{tabular}{cc}
\toprule
 \textbf{Hyperparameter} & \textbf{Value} \\
\midrule
Batch Size & 2 \\
Number of steps for DeepCell database & 2500 {(approx. 38mins)} \\ 
Number of steps for CTC datasets & 700 {(approx. 20mins)} \\ 
Learning rate scheduler & OneCycleLR \\
Upper learning rate & 5e-4 \\
Weight decay & 1e-5 \\
Epsilon & 1e-8 \\
Optimizer & AdamW \\ 
Percentage of the cycle increasing the learning rate & 0.05 \\
Cycle momentum & False\\
Annealing strategy & linear \\
Value of the discount factor $\gamma$ for $\mathcal{L}_{tra}$ & 0.8\\
\bottomrule
\end{tabular}
}}
\caption{\textbf{Optimizer Configuration for CAP.} }
\label{hyperparameters}
\end{table*}

As for the \textbf{testing (evaluation)} process, we preprocess the first frame for each sequence using~\cite{scherr2020cell} to select query points $Q$ and start the process of window rolling.
The evaluation sequence $S_\text{infer}$ follows the native resolution ($384 \times 512$), which the dataset is trained on.
{We set $l_\text{win}=50$ for DeepCell database and $l_\text{win}=100$ for CTC datasets, which is mentioned in Sec.~\ref{infer}.}
We follow the $\mathrm{TRA}$ evaluation metrics~\cite{matula2015cell,kan2013measures} of the operations provided by ISBI~\cite{mavska2023cell,ulman2017objective} and CTMC~\cite{anjum2020ctmc} to correct the situations of merged cells, missing links, wrong links, links with wrong semantics, false positives, and false negatives: $\mathrm{EdgeSplit}$ ($\mathrm{ES}$), $\mathrm{EdgeAdd}$ ($\mathrm{EA}$), $\mathrm{EdgeDelete}$ ($\mathrm{ED}$), $\mathrm{EdgeSemantic}$ ($\mathrm{ESM}$), $\mathrm{FP}$, and $\mathrm{FN}$.
$\mathrm{TRA}$ lies in the range $[0, 1]$ where a higher score corresponds with a better performance, and is calculated by:
\begin{align}
&\mathrm{TRA} = 1-\frac{\mathrm{min}(AOGM, AOGM_0)}{AOGM_0}, \text{ where}\\
&AOGM = w_{\mathrm{ES}}\mathrm{ES} + w_{\mathrm{EA}}\mathrm{EA} + w_{\mathrm{ED}}\mathrm{ED} +
w_{\mathrm{ESM}}\mathrm{ESM} + w_{\mathrm{FP}}\mathrm{FP} + w_{\mathrm{FN}}\mathrm{FN}.
\end{align}
Here $AOGM$ is the weighted sum of the executed operations considered as the cost of transforming ($\mathrm{ES}$, $\mathrm{EA}$, $\mathrm{ED}$, $\mathrm{ESM}$, $\mathrm{FP}$, and $\mathrm{FN}$) the computed graph into the reference one, and $AOGM_0$ is the $AOGM$ value for empty tracking results.
The details of $\mathrm{ES}$, $\mathrm{EA}$, $\mathrm{ED}$, $\mathrm{ESM}$, $\mathrm{FP}$, and $\mathrm{FN}$ are elaborated as follows\footnote{\url{https://celltrackingchallenge.net/evaluation-methodology/}}:

\noindent $\mathrm{\mathbf{ES}}$ is the operation for non-split vertices that the computed markers with more than one reference marker assigned;

\noindent $\mathrm{\mathbf{EA}}$ is the operation for missing links that the cell lineage relations are not indicated;

\noindent $\mathrm{\mathbf{ED}}$ is the operation for redundant links that the fake cell lineage relations are indicated;

\noindent $\mathrm{\mathbf{ESM}}$ is the operation for altering the edge semantics, that is, the difference in semantics between the reference graph and the induced subgraph;

\noindent $\mathrm{\mathbf{FP}}$ is the situation in the extra detected objects;

\noindent $\mathrm{\mathbf{FN}}$ is the situation in the missed objects.

\paragraph{Baselines}
To evaluate the performance of CAP, we conducted a comparison experiment with baseline methods, cell-tracker-gnn (GNN)~\cite{ben2022graph}, EmbedTrack~\cite{embedTrack},  KIT-GE~\cite{scherr2020cell}, Trackastra~\cite{gallusser2024trackastra}, Cell-TRACTR~\cite{o2025cell}, and EBB~\cite{kirsten2025cell}, which have publicly available code and utilize the evaluation system provided by ISBI~\cite{mavska2023cell,ulman2017objective} and the acyclic oriented graphs matching (AOGM) measure~\cite{matula2015cell}.
For the DeepCell data, we follow the experiments~\cite{moen2019accurate} and the default training/testing subsets to report the result.
For the ISBI CTC dataset, to control for variables and ensure fairness, we train these methods on sequence $02$ and evaluate on $01$, following~\cite{scherr2020cell}.
{Since the EBB baseline method does not provide pretrained weights for detection (U373 and GOWT1), we re-run its provided code under the same experimental setup for training, test by the same metrics, and report the results.}
We report the measures of metrics involved in the cell tracking, $\mathrm{EdgeSplit}$, $\mathrm{EdgeAdd}$, $\mathrm{EdgeDelete}$, $\mathrm{EdgeSemantic}$, $\mathrm{FP}$, $\mathrm{FN}$, $\mathrm{TRA}$, and the average inference time ($\mathrm{IT}$) to evaluate the cell tracking performance and the inference efficiency.

\subsection{Results and Analysis}
\label{result_analysis}

We performed comparison experiments of our method, CAP, against baseline approaches, including GNN, EmbedTrack, KIT-GE, Trackastra, Cell-TRACTR, and EBB across the DeepCell database (HeLa, PC-3, 3T3, and RAW264) and the ISBI datasets (U373, GOWT1, and Huh7). The evaluation is based on both cell-tracking accuracy-related metrics ($\mathrm{TRA}$ derived from $\mathrm{EdgeSplit}$, $\mathrm{EdgeAdd}$, $\mathrm{EdgeDelete}$, $\mathrm{EdgeSemantic}$, $\mathrm{FP}$, and $\mathrm{FN}$) and inference efficiency ($\mathrm{IT}$).

\begin{table*}
\centering
\resizebox{\textwidth}{!}{%
\begin{tabular}{llcccccccc}
\toprule
\multirow{2}[2]{*}{Method} & \multirow{2}[2]{*}{Type} & \multicolumn{7}{c}{HeLa} \\
\cmidrule(lr){3-10}
 & & $\mathrm{EdgeSplit}$ $\downarrow$ & $\mathrm{EdgeAdd}$ $\downarrow$ & $\mathrm{EdgeDelete}$ $\downarrow$ & $\mathrm{EdgeSemantic}$ $\downarrow$ & $\mathrm{FP}$ $\downarrow$ & $\mathrm{FN}$ $\downarrow$ & $\mathrm{TRA}$ $\uparrow$ & $\mathrm{IT}$ $\downarrow$\\
\midrule
GNN~\cite{ben2022graph} 22'& SEG+TRA &2 & 385 & 1 & 13 & 62& 342& 0.776& 23.3 \\
EmbedTrack~\cite{embedTrack} 22'& SEG+TRA & 0 & 211 & 1 & 1 & 31 & 218 & 0.849 &  41.5 \\
KIT-GE~\cite{scherr2020cell} 20'& SEG+TRA & 6 & 57 & 1 & 17 & 70 & 67 & \underline{0.923} & 14.4 \\
Trackastra~\cite{gallusser2024trackastra} 24'& SEG+TRA & 1 &  240 & 1 & 1 & 124 & 185 & 0.825 & \underline{11.8}\\
Cell-TRACTR~\cite{o2025cell} 25'& SEG+TRA & 6 & 209 & 1 & 1 & 715 & 144 & 0.816 & 30.3\\
CAP (Ours) & TRA & 3 & 82 & 1 & 0 & 60 & 70 & \textbf{0.926} & \textbf{1.3}\\
\bottomrule
\toprule
\multirow{2}[2]{*}{Method} & \multirow{2}[2]{*}{Type} & \multicolumn{7}{c}{PC-3} \\
\cmidrule(lr){3-10}
 & & $\mathrm{EdgeSplit}$ $\downarrow$ & $\mathrm{EdgeAdd}$ $\downarrow$ & $\mathrm{EdgeDelete}$ $\downarrow$ & $\mathrm{EdgeSemantic}$ $\downarrow$ & $\mathrm{FP}$ $\downarrow$ & $\mathrm{FN}$ $\downarrow$ & $\mathrm{TRA}$ $\uparrow$ & $\mathrm{IT}$ $\downarrow$\\
\midrule
GNN~\cite{ben2022graph} 22'& SEG+TRA &2 & 52 & 3 & 12 & 33& 0& 0.868& 18.3 \\
EmbedTrack~\cite{embedTrack} 22'& SEG+TRA & 0 & 43 & 3 & 1 & 0 & 42 & 0.912 & 48.7 \\
KIT-GE~\cite{scherr2020cell} 20'& SEG+TRA & 29 & 56 & 0 & 0 & 13 & 1 & \underline{0.937} & 9.1\\
Trackastra~\cite{gallusser2024trackastra} 24'& SEG+TRA & 2 & 233 & 2 & 1 & 155 & 190 & 0.317 & \underline{8.9}\\
Cell-TRACTR~\cite{o2025cell} 25'& SEG+TRA & 1 & 26 & 1 & 0 & 146 & 8 & 0.913 & 33.7\\
CAP (Ours) & TRA & 1 & 11 & 0 & 0 & 35 & 7 & \textbf{0.952} & \textbf{1.1}\\
\bottomrule
\toprule
\multirow{2}[2]{*}{Method} & \multirow{2}[2]{*}{Type} & \multicolumn{7}{c}{3T3} \\
\cmidrule(lr){3-10}
 & & $\mathrm{EdgeSplit}$ $\downarrow$ & $\mathrm{EdgeAdd}$ $\downarrow$ & $\mathrm{EdgeDelete}$ $\downarrow$ & $\mathrm{EdgeSemantic}$ $\downarrow$ & $\mathrm{FP}$ $\downarrow$ & $\mathrm{FN}$ $\downarrow$ & $\mathrm{TRA}$ $\uparrow$ & $\mathrm{IT}$ $\downarrow$\\
\midrule
GNN~\cite{ben2022graph} 22'& SEG+TRA & 6 & 613 & 16& 3& 100& 9& \textbf{0.857} & 21.7 \\
EmbedTrack~\cite{embedTrack} 22'& SEG+TRA & 2 & 603 & 14 & 4 & 201 & 589 & 0.853 & 74.6  \\
KIT-GE~\cite{scherr2020cell} 20'& SEG+TRA & 128 & 783 & 14 & 10 & 256 & 698 & 0.820 & 19.1 \\
Trackastra~\cite{gallusser2024trackastra} 24'& SEG+TRA & 10 & 2597 & 16 & 21 & 1511 & 1936& 0.572 & \underline{12.0}\\
Cell-TRACTR~\cite{o2025cell} 25'& SEG+TRA & 6 & 757 & 3 & 10 & 11834 & 213 & 0.727 & 65.0\\
CAP (Ours) & TRA & 53 & 849 & 2 & 4 & 189 & 509 & \underline{0.854} & \textbf{2.2}\\
\bottomrule
\toprule
\multirow{2}[2]{*}{Method} & \multirow{2}[2]{*}{Type} & \multicolumn{7}{c}{RAW264} \\
\cmidrule(lr){3-10}
 & & $\mathrm{EdgeSplit}$ $\downarrow$ & $\mathrm{EdgeAdd}$ $\downarrow$ & $\mathrm{EdgeDelete}$ $\downarrow$ & $\mathrm{EdgeSemantic}$ $\downarrow$ & $\mathrm{FP}$ $\downarrow$ & $\mathrm{FN}$ $\downarrow$ & $\mathrm{TRA}$ $\uparrow$ & $\mathrm{IT}$ $\downarrow$\\
\midrule
GNN~\cite{ben2022graph} 22'& SEG+TRA & 186 & 1451 & 89& 21& 239& 1341& 0.801 & \underline{28.5} \\
EmbedTrack~\cite{embedTrack} 22'& SEG+TRA & 68 & 1475 & 74 & 28 & 669 & 1259 & 0.803 & 59.6  \\
KIT-GE~\cite{scherr2020cell} 20'& SEG+TRA & 37 & 585 & 106 & 24 & 718 & 603 & \textbf{0.876} & 96.7 \\
Trackastra~\cite{gallusser2024trackastra} 24'& SEG+TRA & 14 & 4572 & 82 & 46 & 2810 & 3354& 0.490 & 56.3\\
Cell-TRACTR~\cite{o2025cell} 25'& SEG+TRA & 21 & 1346 & 11 & 12 & 16617 & 354 & 0.713 & 51.5\\
CAP (Ours) & TRA & 67 & 1098 & 12 & 17 & 647 & 898 & \underline{0.864} & \textbf{2.8}\\
\bottomrule
\end{tabular}
}
\caption{
\textbf{Comparison Experiments of DeepCell Database.}
We compare CAP to existing cell tracking frameworks available on the HeLa, PC-3, 3T3, and RAW264 datasets in the DeepCell database. {The method \textit{Type} is defined as follows: the method following SEG+TRA (segmentation + tracking style) and TRA (tracking-only (no explicit segmentation) style).}
}
\label{comp_deepcell}
\end{table*}

Table~\ref{comp_deepcell} demonstrates the quantitative comparison results on DeepCell database. 
On HeLa and PC-3, CAP achieves the highest $\mathrm{TRA}$ scores of $0.926$ and $0.952$, surpassing prior works while requiring only $1.3$s and $1.1$s inference time, respectively. 
This highlights CAP’s ability to deliver state-of-the-art accuracy with minimal computational cost. 
On RAW264, CAP attains a competitive $\mathrm{TRA}$ of $0.864$ with just $2.8$s inference, demonstrating a $3-20\times$ reduction in runtime compared to other methods. 
Even on the challenging 3T3 dataset, CAP achieves a strong $\mathrm{TRA}$ of $0.854$, comparable to the best baseline ($0.857$), but with an inference time of only $2.2$s versus up to $65.0$s for competing methods.
As for the ISBI CTC datasets, as shown in Table~\ref{comp_isbi}, on U373, CAP achieves the best overall performance, recording zero errors across tracking structural metrics and achieving a $\mathrm{TRA}$ of $0.985$, with the lowest inference time of $2.6$s. 
For GOWT1, CAP achieves a $\mathrm{TRA}$ of $0.960$, slightly below KIT-GE ($0.966$), while reducing inference time to $7.1$s, corresponding to a $24\times$ speedup.
On Huh7, CAP matches the best reported $\mathrm{TRA}$ ($0.966$) with the shortest inference time of $3.2$s.

{
Beyond the aggregate TRA score, we further interpret CAP’s performance on cell division/lineage reconstruction using the structural error decomposition that underlies $\mathrm{TRA}$ ($\mathrm{EdgeSplit}$, $\mathrm{EdgeAdd}$, $\mathrm{EdgeDelete}$, and $\mathrm{EdgeSemantic}$). On DeepCell database, CAP yields zero $\mathrm{EdgeSemantic}$ errors on HeLa and PC-3 while maintaining strong $\mathrm{TRA}$, indicating accurate mother–daughter relation recognition rather than only improved runtime. This advantage is especially meaningful on division-scarce PC-3 (only $5$ divisions), where CAP reduces $\mathrm{EdgeAdd}$ to $11$ (vs. $26–233$ for other methods) while keeping $\mathrm{EdgeDelete/EdgeSemantic}$ at $0$, which is consistent with the goal of mitigating division imbalance via AEG sampling. 
On division-rich RAW264 ($1322$ divisions), CAP remains competitive in lineage-structure errors (e.g., $\mathrm{EdgeDelete = 12}$ and $\mathrm{EdgeSemantic = 17}$) while achieving substantially the fastest inference, suggesting robust division tracking under frequent mitosis and crowded conditions. 
Similar trends are observed on the ISBI CTC datasets: CAP achieves all zero structural errors on U373 and maintains low $\mathrm{EdgeSemantic/EdgeDelete}$ on GOWT1 and Huh7. Qualitative lineage examples in Figures~\ref{vis} and~\ref{division_vis} further illustrate CAP’s ability to correctly localize division points and preserve mother-daughter continuity even with simultaneous divisions.}

As for the extreme instability of the EmbedTrack and EBB methods on Huh7 with a $\mathrm{TRA}$ of $0.187$ and $0.626$, respectively, our analysis suggests that the observed results stem from the method’s excessive reliance on the segmentation (SEG) stage and the usage of refined masks during the training of the segmentation model, while Huh7 contains only SEG golden truth (GT) masks without the more refined SEG silver truth (ST).
This further substantiates an end-to-end one-stage cell tracking framework, which bypasses the segmentation or detection stage, enhances model stability, and raises the overall performance limits achievable by the model.

\begin{table*}
\centering
\resizebox{\textwidth}{!}{%
\begin{tabular}{llcccccccc}
\toprule
\multirow{2}[2]{*}{Method} & \multirow{2}[2]{*}{Type} & \multicolumn{7}{c}{U373} \\
\cmidrule(lr){3-10}
 & & $\mathrm{EdgeSplit}$ $\downarrow$ & $\mathrm{EdgeAdd}$ $\downarrow$ & $\mathrm{EdgeDelete}$ $\downarrow$ & $\mathrm{EdgeSemantic}$ $\downarrow$ & $\mathrm{FP}$ $\downarrow$ & $\mathrm{FN}$ $\downarrow$ & $\mathrm{TRA}$ $\uparrow$ & $\mathrm{IT}$ $\downarrow$\\
\midrule
GNN~\cite{ben2022graph} 22'& SEG+TRA &214 & 371 & 0 & 0 & 135& 25& 0.771& \underline{19.7} \\
EmbedTrack~\cite{embedTrack} 22'& SEG+TRA & 0 & 44 & 0 & 7 & 211 & 32 & 0.931 &  123.6 \\
KIT-GE~\cite{scherr2020cell} 20'& SEG+TRA & 3 & 50 & 0 & 1 & 197 & 25 & 0.939 & 22.0 \\
Trackastra~\cite{gallusser2024trackastra} 24'& SEG+TRA & 3 & 50 & 0 & 21 & 182 & 25 & 0.938 & 20.3\\
 EBB~\cite{kirsten2025cell} 25'& DET+TRA &  2 & 37 & 0 & 22 & 143 & 22 &\underline{0.959} & 45.6\\
CAP (Ours) & TRA & 0 & 0 & 0 & 0 & 119 & 0 & \textbf{0.985} & \textbf{2.6}\\
\bottomrule
\toprule
\multirow{2}[2]{*}{Method} & \multirow{2}[2]{*}{Type} & \multicolumn{7}{c}{GOWT1} \\
\cmidrule(lr){3-10}
 & & $\mathrm{EdgeSplit}$ $\downarrow$ & $\mathrm{EdgeAdd}$ $\downarrow$ & $\mathrm{EdgeDelete}$ $\downarrow$ & $\mathrm{EdgeSemantic}$ $\downarrow$ & $\mathrm{FP}$ $\downarrow$ & $\mathrm{FN}$ $\downarrow$ & $\mathrm{TRA}$ $\uparrow$ & $\mathrm{IT}$ $\downarrow$\\
\midrule
GNN~\cite{ben2022graph} 22'& SEG+TRA &49 & 140 & 1 & 13 & 5& 28& 0.958& 67.2 \\
EmbedTrack~\cite{embedTrack} 22'& SEG+TRA & 39 & 119 & 0 & 0 & 21 & 40 & 0.961 & 171.6 \\
KIT-GE~\cite{scherr2020cell} 20'& SEG+TRA & 45 & 111 & 0 & 1 & 6 & 32 & \underline{0.966} & 39.4\\
Trackastra~\cite{gallusser2024trackastra} 24'& SEG+TRA & 47 & 140 & 0 & 1 & 4 & 36 & 0.965 & \underline{27.6}\\
 EBB~\cite{kirsten2025cell} 25'& DET+TRA & 45 & 129 & 0 & 0 & 5 & 34 & \textbf{0.980} & 50.3\\
CAP (Ours) & TRA & 18 & 120 & 2 & 2 & 75 & 46 & 0.960 & \textbf{7.1}\\
\bottomrule
\toprule
\multirow{2}[2]{*}{Method} & \multirow{2}[2]{*}{Type} & \multicolumn{7}{c}{Huh7} \\
\cmidrule(lr){3-10}
 & & $\mathrm{EdgeSplit}$ $\downarrow$ & $\mathrm{EdgeAdd}$ $\downarrow$ & $\mathrm{EdgeDelete}$ $\downarrow$ & $\mathrm{EdgeSemantic}$ $\downarrow$ & $\mathrm{FP}$ $\downarrow$ & $\mathrm{FN}$ $\downarrow$ & $\mathrm{TRA}$ $\uparrow$ & $\mathrm{IT}$ $\downarrow$\\
\midrule
GNN~\cite{ben2022graph} 22'& SEG+TRA & 34 & 85 & 2& 835& 36& 4& 0.889 & 27.1 \\
EmbedTrack~\cite{embedTrack} 22'& SEG+TRA & 0 & 733 & 0 & 5 & 1392 & 636 & 0.187 & 125.0  \\
KIT-GE~\cite{scherr2020cell} 20'& SEG+TRA & 34 & 77 & 0 & 0 & 41 & 0 & \textbf{0.970} & 25.7 \\
Trackastra~\cite{gallusser2024trackastra} 24'& SEG+TRA & 34 & 84 & 2 & 5 & 34 & 5& 0.964 & \underline{18.2}\\
 EBB~\cite{kirsten2025cell} 25'& DET+TRA & 27 & 588 & 0 & 4 & 870 & 247 & 0.620 &  19.0\\
CAP (Ours) & TRA & 31 & 71 & 0 & 1 & 101 & 1 & \underline{0.966} & \textbf{3.2}\\
\bottomrule
\end{tabular}
}
\caption{
\textbf{Comparison Experiments of the ISBI CTC Dataset.}
We compare CAP to existing cell tracking frameworks available on the datasets PhC-C2DH-U373, Fluo-N2DH-GOWT1, and Fluo-C2DL-Huh7. {The method \textit{Type} is defined as follows: the method following SEG/DET+TRA (segmentation/detection + tracking style) and TRA (tracking-only (no explicit segmentation/detection stage) style).}
}
\label{comp_isbi}
\end{table*}

\begin{figure*}[t]
\centering
\centerline{\includegraphics[width=\textwidth]{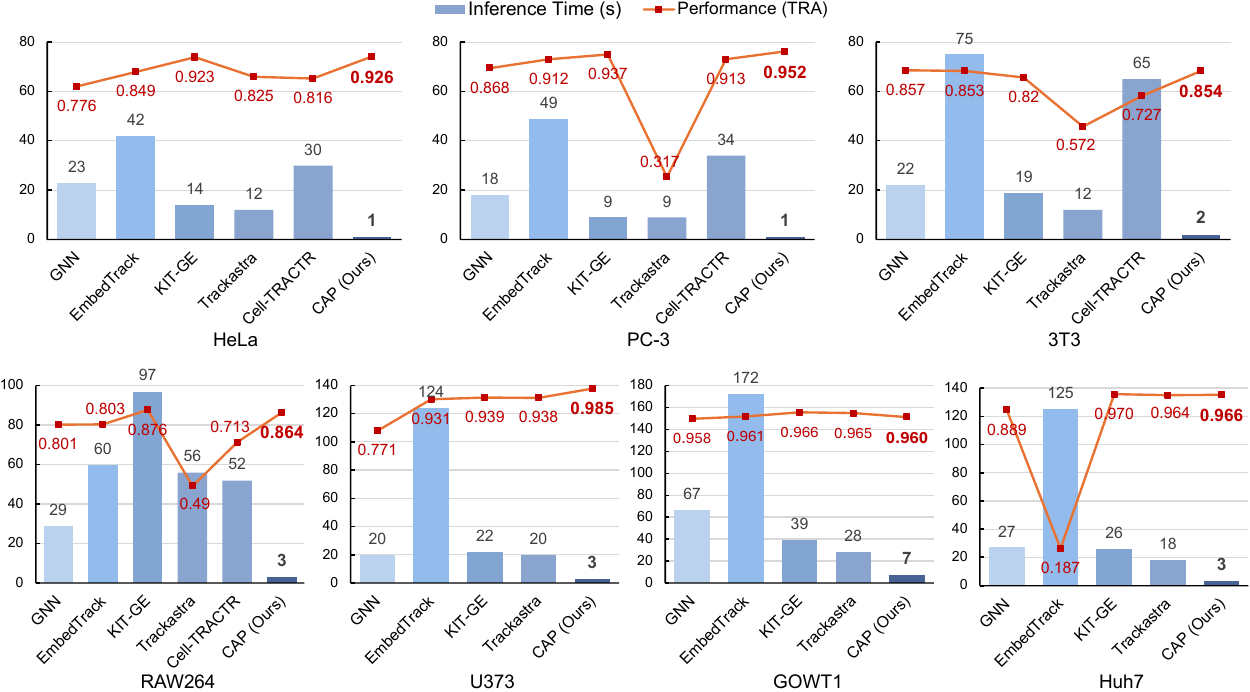}}
\caption{\textbf{Result of Tracking Accuracy and Inference Efficiency across Multiple Benchmark Datasets.} CAP consistently achieves the best trade-off between accuracy and efficiency, delivering competitive or superior performance while reducing inference time.
}
\label{charts}
\end{figure*}

Overall, as shown in Figure~\ref{charts}, the CAP method demonstrates promising performance and constant high efficiency on inference and label usage across all datasets, consistently achieving high $\mathrm{TRA}$ scores and the lowest $\mathrm{IT}$s.
Thus, CAP is a highly effective, efficient, and reliable method for cell tracking in a practical biomedical field.

\subsection{Qualitative Result}

\begin{figure*}[t]
\centering
\centerline{\includegraphics[width=\textwidth]{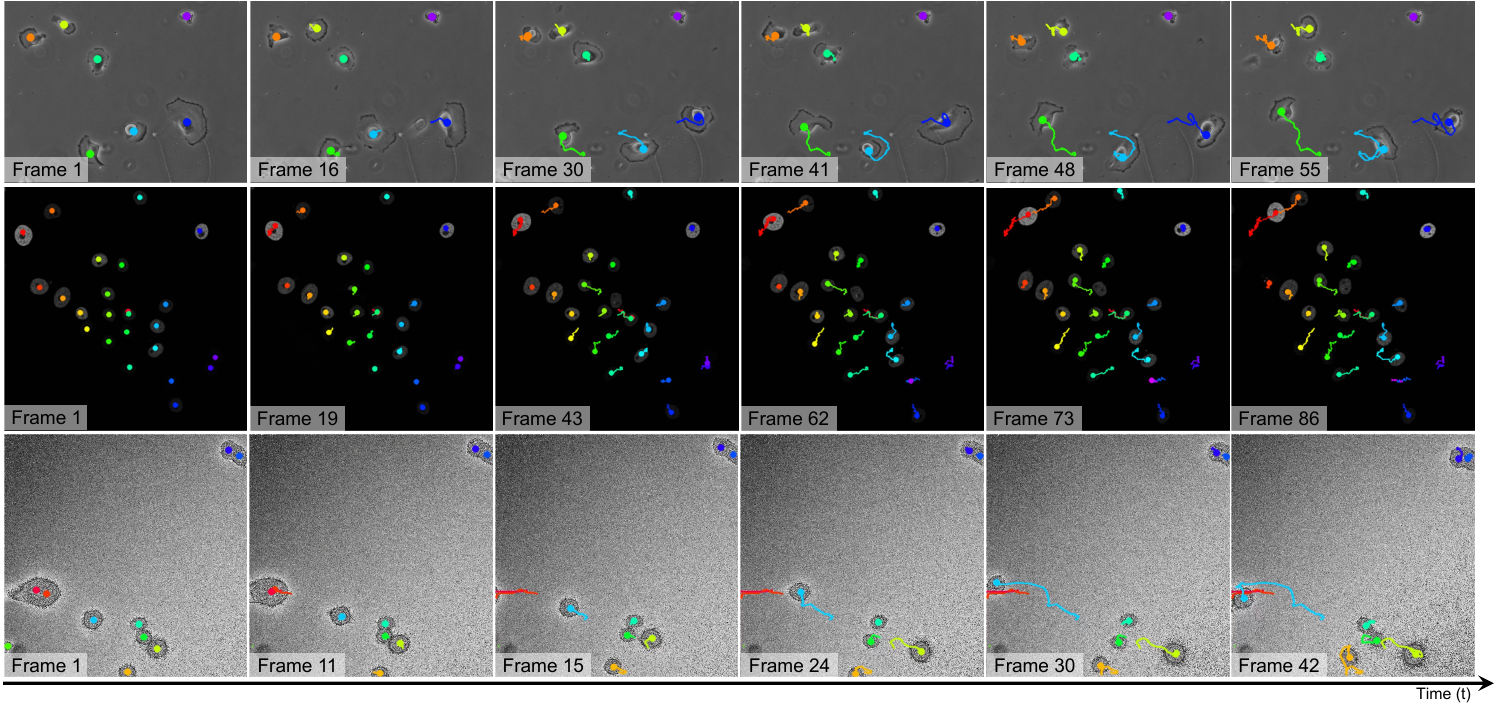}}
\caption{\textbf{Cell Trajectories.} Each point represents a cell with distinct colors, indicating individual cell trajectories across time points. The cell density for each trajectory follows the order: medium, high, and low.}
\label{vis}
\end{figure*}

To further validate the performance of CAP, we provide qualitative results illustrating two core aspects of cell tracking: cell trajectory estimation and lineage (division) reconstruction.

Figure~\ref{vis} visualizes representative examples of cell trajectories across multiple time points. 
Each cell is assigned a distinct color, enabling clear visualization of its movement path. The results demonstrate CAP’s ability to robustly maintain cell identities over extended sequences, even in the presence of densely packed regions, complex motion patterns, and various motion speeds. This consistency highlights CAP’s strength in long-term trajectory preservation and cross-frame association accuracy.

\begin{figure}[t]
\centering
\centerline{\includegraphics[width=\textwidth]{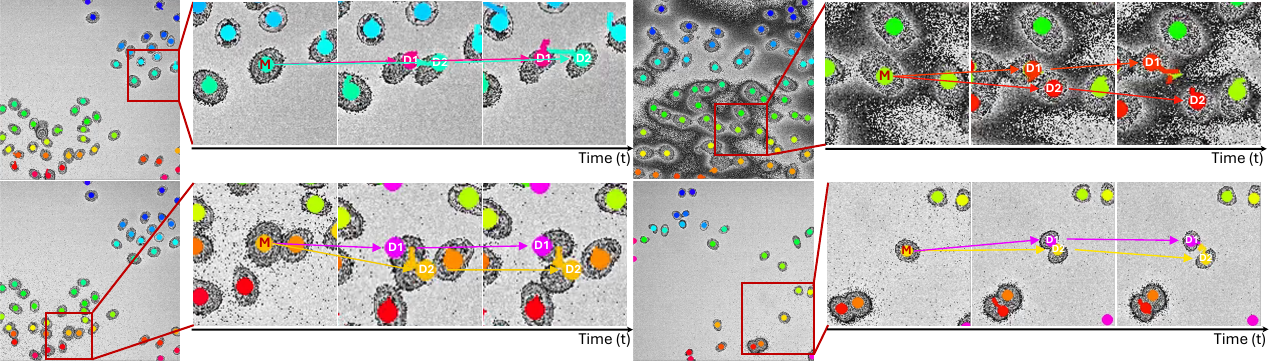}}
\caption{\textbf{Cell Lineage (Divisions).} Utilizing the proposed Cell Point Trajectory and Visibility representation illustrated in Figure~\ref{fig:cell_represent}, CAP is able to correctly identify the cell division and track all cells in the sequence. \textbf{M} denotes the mother cell; \textbf{D1} and \textbf{D2} represent the daughter cell \#1 and the daughter cell \#2.
}
\label{division_vis}
\end{figure}

Figure~\ref{division_vis} illustrates the model’s performance on capturing cell division events. 
CAP successfully identifies division points and accurately tracks the transition from mother cells (M) to daughter cells (D1 and D2), ensuring continuity across frames. 
The highlighted examples show that CAP not only maintains the mother–daughter relationships but also effectively disambiguates multiple simultaneous divisions within crowded environments. 
These results underscore the model’s capacity to reconstruct biologically meaningful lineage information, which is critical for downstream analysis in developmental biology and cancer research.

\subsection{Ablation Study}
\label{ablation}
We performed extensive experiments to assess the effectiveness of each key component in the CAP framework on both the DeepCell and ISBI CTC datasets: the AEG sampling strategy, the cell joint tracking structure, and the feature stride of the model.
To more clearly observe the influence of each key component on individual test sequences, we report the $\mathrm{TRA}$ results for each sequence in the DeepCell database.

\begin{table*}
\centering
\resizebox{\textwidth}{!}{%
\setlength{\tabcolsep}{1mm}{
\begin{tabular}{ccccccccccccccc}
\toprule
\multirow{2}[2]{*}{AEG Sampling} & \multirow{2}[2]{*}{U373}& \multirow{2}[2]{*}{GOWT1} & \multirow{2}[2]{*}{Huh7} & \multicolumn{5}{c}{HeLa}  & \multicolumn{2}{c}{PC-3}  & \multicolumn{1}{c}{3T3}  & \multicolumn{3}{c}{RAW264}\\
\cmidrule(lr){5-9}
\cmidrule(lr){10-11}
\cmidrule(lr){12-12}
\cmidrule(lr){13-15}
 &&&& 01 & 02 & 03 & 04 &05& 01 & 02& 01 & 01& 02& 03\\
\midrule
\ding{55} & 0.556  & 0.941  & 0.942 & 0.898  & 0.908  & 0.883 & \textbf{0.921} & 0.878 & 0.952 & 0.924 & 0.829 & 0.804 & 0.828 & 0.843\\
\ding{51} & \textbf{0.985} & \textbf{0.960} & \textbf{0.966} & \textbf{0.962} & \textbf{0.958} & \textbf{0.921} & 0.900 & \textbf{0.891} & \textbf{0.980} & \textbf{0.924} & \textbf{0.832} & \textbf{0.824}&\textbf{0.868} & \textbf{0.900}\\
\bottomrule
\end{tabular}
}}
\caption{\textbf{Effectiveness of AEG Sampling.} AEG sampling is built to balance the cell event (division) in a training sequence.}
\label{tab:abl_aeg}
\end{table*}

\paragraph{Effectiveness of AEG Sampling}

To address the imbalance of cell division events during training, we propose the Adaptive Event-Guided (AEG) sampling strategy (Section~\ref{aeg}), which calibrates the training process by ensuring balanced exposure to complete division sequences. 
As shown in Table~\ref{tab:abl_aeg}, disabling AEG sampling leads to substantial performance degradation across datasets. 
With AEG sampling enabled, the model achieves consistent improvements: for instance, on the HeLa dataset, the best $\mathrm{TRA}$ score rises from $0.898$ to $0.962$, while on U373 it improves from $0.556$ to $0.985$. Similar gains are observed on PC-3 and RAW264. These results confirm that AEG sampling effectively balances event occurrences and significantly enhances the model’s ability to learn from imbalanced data.

\begin{table*}
\centering
\resizebox{\linewidth}{!}{
\setlength{\tabcolsep}{1mm}{
\begin{tabular}{cccccccccccccccc}
\toprule
\multicolumn{2}{c}{Attention} & \multirow{2}[2]{*}{U373}& \multirow{2}[2]{*}{GOWT1} & \multirow{2}[2]{*}{Huh7} & \multicolumn{5}{c}{HeLa}  & \multicolumn{2}{c}{PC-3}  & \multicolumn{1}{c}{3T3}  & \multicolumn{3}{c}{RAW264}\\
\cmidrule(lr){1-2}
\cmidrule(lr){6-10}
\cmidrule(lr){11-12}
\cmidrule(lr){13-13}
\cmidrule(lr){14-16}
time & joint  & && &01 & 02 & 03 & 04 &05& 01 & 02& 01 & 01& 02& 03\\
\midrule
\ding{51}& \ding{55} & 0.974 &  0.903  & 0.944 & 0.853  & 0.949  & 0.861 & \textbf{0.921} & \textbf{0.906} & 0.925 & \textbf{0.947} & 0.812 & 0.805 & 0.845 & 0.821\\
\ding{51}& \ding{51} & \textbf{0.985} & \textbf{0.960} & \textbf{0.966} & \textbf{0.962} & \textbf{0.958} & \textbf{0.921} & 0.900 & 0.891 & \textbf{0.980} & 0.924 & \textbf{0.832} & \textbf{0.824}&\textbf{0.868} & \textbf{0.900}\\

\bottomrule
\end{tabular}
}}
\caption{\textbf{Effectiveness of Cell Joint Tracking.}
We compare CAP performance using time and cross-trajectory attention.}
\label{tab:joint}
\end{table*}

\paragraph{Effectiveness of Cell Joint Tracking}
We evaluate the role of joint tracking by replacing cross-trajectory attention layers with time-only attention while maintaining model size. 
As reported in Table~\ref{tab:joint}, this modification consistently reduces performance, with $\mathrm{TRA}$ drops ranging from $0.015$ to $0.055$ across datasets. Incorporating both time and joint attention yields notable improvements, particularly on HeLa where the $\mathrm{TRA}$ increases from $0.853$ to $0.962$ in the best case. This demonstrates that the dual-attention design not only strengthens temporal coherence but also captures relational dependencies between trajectories, leading to more robust tracking in complex cellular environments.

\begin{table}[t]
\resizebox{\linewidth}{!}{
\setlength{\tabcolsep}{1mm}{
\begin{tabular}{ccccccccccccccc}
\toprule
\multirow{2}[2]{*}{Stride} & \multirow{2}[2]{*}{U373}& \multirow{2}[2]{*}{GOWT1} & \multirow{2}[2]{*}{Huh7} & \multicolumn{5}{c}{HeLa}  & \multicolumn{2}{c}{PC-3}  & \multicolumn{1}{c}{3T3}  & \multicolumn{3}{c}{RAW264} \\
\cmidrule(lr){5-9}
\cmidrule(lr){10-11}
\cmidrule(lr){12-12}
\cmidrule(lr){13-15}
 &&&& 01 & 02 & 03 & 04 &05& 01 & 02& 01 & 01& 02& 03\\
\midrule
8 & 0.531 & 0.609  & 0.853 & 0.884  & 0.919  & 0.806 & \textbf{0.936} & 0.853 & 0.105 & 0.274 & 0.387 & 0.610 & 0.622 & 0.758\\
4 & \textbf{0.985} & \textbf{0.960} & \textbf{0.966} & \textbf{0.962} & \textbf{0.958} & \textbf{0.921} & 0.900 & \textbf{0.891} & \textbf{0.980} & \textbf{0.924} & \textbf{0.832} & \textbf{0.824}&\textbf{0.868} & \textbf{0.900}\\
\bottomrule
\end{tabular}
}}
\caption{\textbf{Ablation on Feature Stride.}
CAP predicts more accurately with higher resolution features.
}

\label{tab:stride}
\end{table}

\paragraph{Ablation on Feature Stride}
Table~\ref{tab:stride} highlights the effect of feature stride on model accuracy. 
Reducing the stride from $8$ to $4$ substantially improves performance by providing higher-resolution features. 
For example, on the HeLa dataset, the best $\mathrm{TRA}$ score increases from $0.884$ to $0.962$, while on U373 the score rises from $0.531$ to $0.985$. 
Similar trends are observed across 3T3 and RAW264, demonstrating that higher-resolution features are critical for capturing fine-grained cell dynamics and improving tracking accuracy, especially in challenging datasets with small or densely packed cells. 
Consequently, we adopt $s=4$ as the default stride setting in the CAP framework.

\begin{table}[t]
\centering
\setlength{\tabcolsep}{2mm}
{
\begin{subtable}[t]{0.43\linewidth}
\centering
\resizebox{\linewidth}{!}{%
\begin{tabular}{c c c c}
\toprule
$l_{win}$ & U373 & GOWT1 & Huh7 \\
\midrule
90 & 0.957 / 2.9 & 0.946 / 7.7 & \textbf{0.967} / 3.7\\
\textbf{100} & \textbf{0.985} / 2.6 & \textbf{0.960} / \textbf{7.1} & 0.966 / 3.2\\
110 & 0.983 / \textbf{2.5} & 0.958 / 7.2 & 0.966 / \textbf{3.1}\\
\bottomrule
\end{tabular}%
}
\end{subtable}}
\hfill
\begin{subtable}[t]{0.55\linewidth}
\centering
\resizebox{\linewidth}{!}{%
{
\begin{tabular}{c cc c ccc}
\toprule
$l_{win}$ & HeLa & PC-3 & 3T3 &RAW264 \\
\midrule
40  & 0.921 / 1.9 & 0.938 / 1.6 & 0.836 / 3.5 & \textbf{0.866} / 3.1 \\
\textbf{50}  & \textbf{0.926} / \textbf{1.3} & \textbf{0.952} / \textbf{1.1} & 0.854 / \textbf{2.2} & 0.864 / \textbf{2.8} \\
60  & 0.923 / 1.5 & 0.950 / 1.4 & \textbf{0.856} / \textbf{2.2} & 0.859 / 3.0 \\
\bottomrule
\end{tabular}%
}}
\end{subtable}
\caption{\textbf{Ablation on Window Size in RAW.} The results indicate TRA values with inference times ($\mathrm{TRA / IT}$). Setting $l_{win} = 50$ for the DeepCell database and $l_{win} = 100$ for the ISBI CTC datasets during RAW inference.}
\label{tab:stride_split}
\end{table}

\paragraph{{Ablation on Window Size}}
Table~\ref{tab:stride_split} reports an ablation of the RAW rolling window size $l_{win}$, showing $\mathrm{TRA}$ together with the inference time (IT). Across the ISBI CTC datasets, $l_{win}=100$ provides the best overall trade-off: it yields the highest TRA on U373 and GOWT1 while also achieving the fastest or near-fastest inference ($2.6$s and $7.1$s, respectively), and it remains essentially tied on Huh7 at $l_{win}=90$. On the DeepCell database, $l_{win}=50$ similarly offers the most favorable balance, delivering the top $\mathrm{TRA}$ on HeLa and PC-3 with the lowest inference times, while matching or nearly matching the best results on 3T3 with comparable IT and RAW264. Based on these observations, we set $l_{win}=100$ for ISBI CTC and $l_{win}=50$ for DeepCell in all experiments.

\section{Conclusion and Future Work}
\label{sec:conclusion}
In this paper, we introduce the one-stage CAP (Cell as Point) framework, which eliminates the need for a separate segmentation or detection stage, a significant innovation in cell tracking. 
This approach substantially improves efficiency via reducing inference time by a factor of $8$ to $32$ and minimizing the need for high-quality data annotation. 
CAP effectively captures complex cell states and tracks cells in one stage while maintaining high efficiency. 
Additionally, it addresses common challenges in cell sequence data, such as cell event imbalance and long sequence inference. 
In conclusion, CAP not only offers considerable practical value in real-world applications due to its efficiency but also pioneers a novel approach in the field of cell tracking.

\paragraph{Limitations and Future Work}
The limitation of our study arises from the nature of the datasets used for evaluation. These datasets were annotated by different experts, and variations in annotation practices may lead to differences in the location of the cell points that CAP generated. 
Such inconsistencies can introduce discrepancies in performance across datasets, as the model could be sensitive to variations in label quality and annotation style. In addition, differences in imaging devices and cell densities across datasets may further amplify these effects, contributing to dataset-specific performance variation.
Consequently, while CAP demonstrates strong trade-offs between accuracy and efficiency, improving generalization remains important for practical deployment. In future work, we will target generalization across: (i) cross-lab/cross-microscope domain shifts, including different sensors, optics, staining/fluorescence conditions, and illumination changes; (ii) cell populations with substantially different morphology and motion statistics, such as highly deformable cells or fast-moving immune cells; (iii) challenging crowding regimes, including very high cell density, frequent occlusions, and long-term interactions where identity switches are common; and (iv) varying acquisition protocols, e.g., different frame rates, missing frames, and longer sequences.

\bibliographystyle{elsarticle-num} 
\bibliography{manuscript}

\end{document}